\documentclass[prb,final,groupedaddress,footinbib,floats]{revtex4}
\usepackage{amsmath}
\usepackage{graphicx}
\usepackage{rotating}
\newcommand{\ts}{{{\sigma\nu z}}}

\begin{document}
\bibliographystyle{apsrev}

\title{Noise in disordered systems: The power spectrum and dynamic
exponents in avalanche models}

\author{Matthew C. Kuntz}
\email[]{mck10@cornell.edu}
\author{James P. Sethna}
\email[]{sethna@lassp.cornell.edu}
\affiliation{LASSP, Department of Physics, Cornell University, Ithaca,
NY 14853}

\date{\today}

\begin{abstract}
For a long time, it has been known that the power spectrum of
Barkhausen noise had a power-law decay at high frequencies.  Up to
now, the theoretical predictions for this decay have been incorrect,
or have only applied to a small set of models.  In this paper, we
describe a careful derivation of the power spectrum exponent in
avalanche models, and in particular, in variations of the
zero-temperature random-field Ising model.  We find that the naive
exponent, $(3-\tau)/\ts$, which has been derived in several other
papers, is in general incorrect for small $\tau$, when large
avalanches are common.  ($\tau$ is the exponent describing the
distribution of avalanche sizes, and $\ts$ is the exponent describing
the relationship between avalanche size and avalanche duration.)  
We find that for a large class of avalanche models, including several
models of Barkhausen noise, the correct exponent for $\tau<2$ is
$1/\ts$.  We explicitly derive the mean-field exponent of $2$. In the 
process, we calculate the average avalanche shape for avalanches of fixed
duration and scaling forms for a number of physical properties.
\end{abstract}

\maketitle

\section{Introduction}
Many physical systems, from superconductors\cite{sc} to
sand-piles\cite{sand_piles} to martensitic shape-memory
alloys\cite{shape_memory} produce noise with power law
characteristics.  The noise in many of these systems can be modeled in
terms of avalanches.  One example of this is Barkhausen noise in
ferromagnetic materials.  Many ferromagnetic materials magnetize not
smoothly, but in jumps of all sizes.  The resulting ``noise'' is
characterized by power laws.  For example, the distributions of
avalanche sizes, durations and energies are all seen to be power laws.
Recently, it has been proposed that this noise is a result of either a
disorder induced critical
point\cite{sethna1,sethna2,sethna3,sethna4,sethna5,dahmen}, or self
organized criticality (SOC)\cite{urbach,stanley1}.  Several variations
of the zero temperature random field Ising model have been proposed to
model this critical behavior.

One of the main power laws which must be explained if these models are
to be successful is the power law behavior of the power spectrum.
Actually, the power spectrum exhibits two different power laws, one
for low frequencies, and another for high frequencies.  The high
frequency power law, $\mathcal{P}_{\text{av}}(\omega)$, reflects the
dynamics within avalanches, and the low frequency power law,
$\mathcal{P}_{\text{corr}}(\omega)$, reflects the correlations between
avalanches. Simulations of various random field Ising models have been
fairly successful in modeling the high frequency scaling of the power
spectrum, but theoretical predictions of this scaling have been absent
or wrong.\cite{lieneweg,sethna2,sethna5,dahmen,stanley2} (See the
discussion in section \ref{ap:prev} for a description of several
previous calculations.)  In this paper, we will derive an exponent
relation for the high frequency power spectrum which applies to
several variations of the random-field Ising model.  Large portions of
the derivation should also apply to any critical avalanche model.

\section{The models}
Several variations of the zero temperature random field Ising model have
been proposed to explain the power laws in Barkhausen noise.  They are
differentiated on the basis of the presence of long range forces, and
the details of the dynamics.  In a separate paper, we examine in
detail the differences between these models\cite{kuntz}.  A general Hamiltonian
for the models is
\begin{align}
\label{eq:hamiltonian}
\mathcal{H} = &-\sum_{\text{nn}} J_{\text{nn}} s_i s_j - \sum_i H s_i -
\sum_i h_i s_i \nonumber \\
	&+ \sum_i \frac{J_{\text{inf}}}{N}\,s_i - \sum_{\{i,j\}}
J_{\text{dipole}}\, \frac{3 \cos (\theta_{ij}) - 1}{r_{ij}^3}\,
s_i s_j,
\end{align}
where $s_i=\pm 1$ is an Ising spin, $J_{\text{nn}}$ is the
strength of the ferromagnetic nearest neighbor interactions, $H$ is an
external magnetic field, $h_i$ is a random local field,
$J_{\text{inf}}$ is the strength of the infinite range demagnetizing
field,\cite{urbach} and $J_{\text{dipole}}$ is the strength of the
dipole-dipole interactions.  The critical exponents of the power laws are
independent of the
particular choice of random field distributions $\rho(h_i)$ for a large
variety of distributions.  Most commonly, a Gaussian distribution of
random fields is used, with a standard deviation $R$.  (When we refer to
the strength of the disorder, we are referring to the width, $R$, of the
random field distribution.)

Two different dynamics have been considered.  The first is a front
propagation dynamics in which spins on the edge of an existing front
flip as soon as it would decrease their energy to do so.  Spins with
no flipped neighbors cannot flip even if it would be energetically
favorable.  Second is a dynamics which includes domain nucleation.
Any spin can flip when it becomes energetically favorable to do so.
In both cases, spins flip in shells---all spins which can flip at time
$t$ flip, then all of their newly flippable neighbors flip at time
$t+1$.

Depending on which terms are included in the Hamiltonian, the behavior
appears to fall into three different universality classes.  When domain
nucleation is allowed, and only nearest neighbor interactions are
included, there is a second order critical point at a particular
disorder, $R_c$, and external field $H_c$.  For disorders below $R_c$, a
finite fraction of the spins in the system (even in the thermodynamic
limit) flip in a single jump.  This critical point seems to have a wide
critical region,\cite{sethna1,sethna2,sethna3,sethna4,sethna5,dahmen} so
this critical behavior might be seen in experiments even without tuning
the disorder. For disorders below $R_c$, or when domain nucleation is
not allowed, the addition of an infinite-range demagnetizing
field\cite{urbach}\cite{NotMeanField}
can self-organize the system to a
different critical behavior.  (Self-organization means that the system
naturally sits at a critical point, without having to tune any
parameters.) The infinite-ranged interaction is sometimes introduced
to mimic the effects of the boundaries of materials with dipolar, or other
long-ranged interactions\cite{stanley1}; these interactions also self-organize 
the model to the critical point\cite{narayan2}. Our use of the term
``infinite-range models'' perhaps
obscures the clear physical origin of this universality class of models.
The critical behavior in front propagation was originally described in a
non-self-organized depinning model by Ji and
Robbins.\cite{robbins}$^,$\cite{BelowRcFrontProp}
Zapperi {\it et al.}\cite{stanley1} claim that the addition of
dipole-dipole interactions to the infinite-range model lowers the upper
critical dimension to three and produces mean-field exponents in three
dimensions.  Since large mean-field simulations are much easier than
large simulations with dipole-dipole interactions, we will give results
from mean-field simulations in this paper.  Dipolar interactions without
an infinite-range term were explored by Magni\cite{magni} in two dimensions,
who found labyrinthine patterns and hysteresis loops similar to those seen in
garnet films.  The relevant exponents for
the three universality classes can be found in table
\ref{tab:exponents}.  Except in section \ref{sec:snz}, all of the
results in this paper are from three-dimensional simulations, disorder $R=1.8$,
nearest neighbor interaction $J_{\rm nn}=1$, and infinite range interaction
$J_{\rm inf}=0.25$; this model exhibits self-organized front-propagation
exponents.

\begin{table}
\begin{center}
\begin{tabular}{lccc}
\hline\hline
		& Short Range (3D) & Front Propagation (3D) 	& Mean
		Field \\
\hline
$\tau$		& 1.6		   & 1.28	& 1.5 \\
$\ts$		& 0.58		   & 0.58	& 0.5 \\
\hline\hline
		& Short Range (4D) & Front Propagation (4D) \\
\hline
$\tau$		& 1.53		& 1.42	\\
$\ts$		& 0.52		& 0.56	\\
\hline\hline
\end{tabular}
\end{center}
\caption{Important exponents for the three universality classes. 
$\tau$ is the exponent for the avalanche size distribution 
$D(S)\sim S^{-\tau}$.  $\ts$ relates the avalanche size $S$ to the 
avalanche duration $T$ : $T \sim S^{\ts}$}
\label{tab:exponents} 
\end{table}

\section{Deriving the exponents}
In the process of deriving the form of the critical exponent for the
power spectrum or energy spectrum (power output during the
duration of the experiment),
we will derive the scaling forms of several other
quantities which are themselves of interest.  Warned by the failure of
the naive scaling exponent for the power spectrum, we will present
numerical scaling plots for each of these intermediate quantities.

\subsection{The avalanche shape}
\begin{figure}
\centerline{\includegraphics[width=0.65\textwidth]{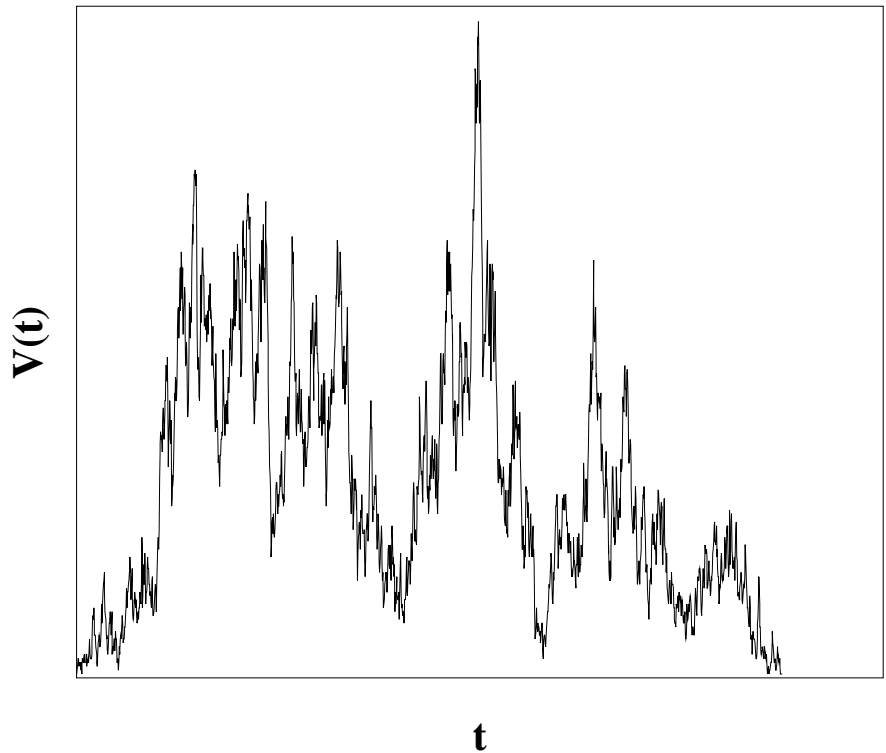}}
\caption{The shape of a typical large avalanche.  Notice that the
avalanche nearly stopped several times, and the voltage (the number of
spins flipped in each time step) fluctuated drastically.}
\label{fig:aval}
\end{figure}

Near criticality, the avalanches in the random-field Ising model have
a very ragged shape.  There are avalanches of all sizes precisely
because each avalanche is always finely balanced between continuing
and dying out.  Most large avalanches come close to dying many times. The
time series for a 
typical large avalanche can be seen in figure \ref{fig:aval}. Petta 
{\it et al.}\cite{petta} have emphasized that the non-trivial structure
within avalanches like these are crucial to understanding the power spectra,
and a good test of the models.

\begin{figure}
\centerline{\includegraphics[width=0.65\textwidth]{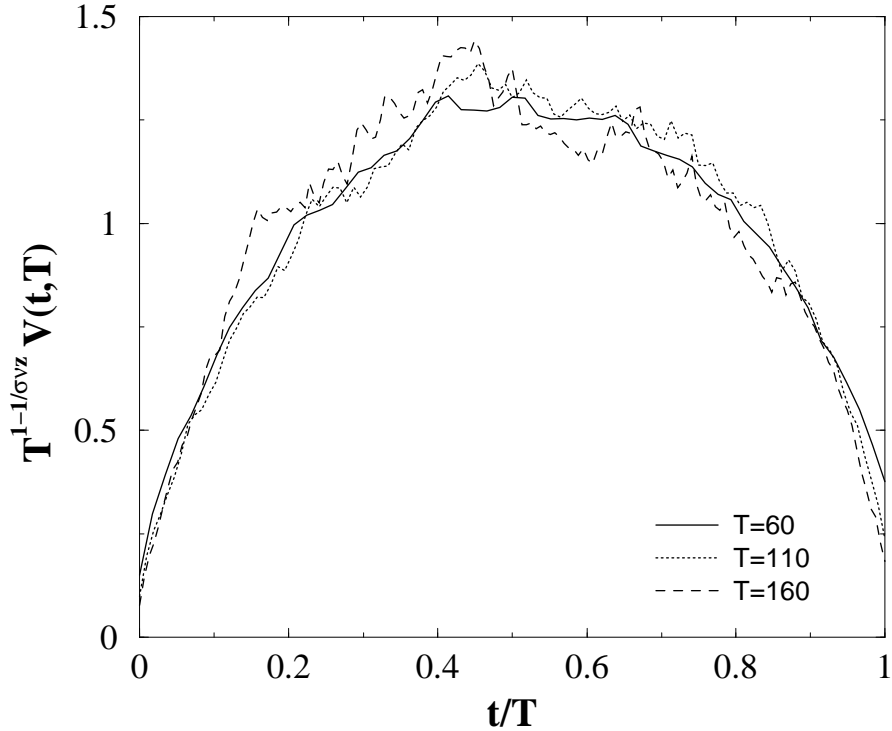}}
\caption{The average avalanche shape for three different durations.
Spasojevi{\'c} {\it et al.}\cite{stanley2} measured the average avalanche
shape experimentally and found a somewhat different shape.  This kind
of measurement provides a much sharper test for the theory than the
tradition of comparing critical exponents.  Presumably, the average
avalanche shape for large sizes and times is a universal scaling
function $f_{\rm shape}$: if the experiment differs in this regard from
our model, then our model is expected to have different critical exponents. All
features at long length and time scales should be universal.}
\label{fig:avg_aval}
\end{figure}

Despite this rough shape, criticality implies that the average
avalanche shape will scale in a universal way.  Consider the average
shape of avalanches of duration $T$.  If we rescale the time axis,
$t$, by a factor of $T$, and divide the vertical axis, which measures
the number of spins flipped (proportionate, in simple cases, to the voltage $V$
which would be
measured in a pickup coil), by the average voltage, we should get a
generic shape which is independent of $T$.  The average height is the
average area, $S(T)\sim T^{1/\ts}$ (let this define the exponent
$\ts$),\cite{sigmanuz}
divided by the duration, $T$.  Therefore, the scaling form
should be
\begin{equation}
V(T,t) = T^{1/\ts - 1} f_{\rm shape}(t/T).
\label{eq:average_aval}
\end{equation}
The scaling of the average avalanche shape according to equation
\ref{eq:average_aval} can be seen in figure
\ref{fig:avg_aval}.\cite{invertedparabola}

\subsection{Distribution of voltages $V$ in avalanches of size $S$}

The power spectrum is sensitive not only to the shapes of the
avalanches, but to the fluctuations in the avalanche shapes.  An
interesting measure of these fluctuations is the probability $P(V|S)$
that a voltage $V$ will occur at some point in an avalanche of size
$S$.  If this probability scales universally, then we know it must
have the form
\begin{equation}
\label{eq:generic_voltage}
P(V|S)=V^{-x} f_{\rm power}(VS^{-y}).
\end{equation}
But what are the exponents $x$ and $y$?  We can determine the value of
$x$ by integrating over all voltages.  Since $P(V|S)$ is a probability
distribution, it must integrate to 1:
\begin{equation}
\int_0^\infty V^{-x} f_{\rm power}(V S^{-y}) dV = A S^{-y(x-1)} = 1.
\label{eq:int_prob}
\end{equation}
From this, we know that $x=1$.  (The alternative, $y=0$, can be
discarded because along with equation \ref{eq:generic_voltage} it
would imply that $P(V|S)$ is independent of the avalanche size $S$.)

\begin{figure}
\centerline{\includegraphics[width=0.65\textwidth]{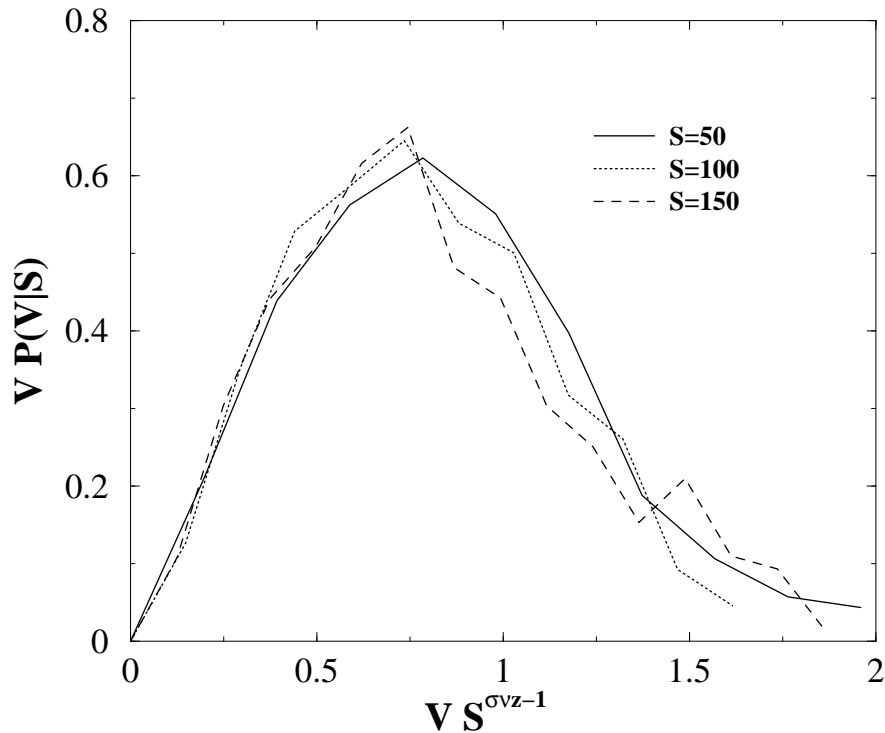}}
\caption{A collapse of the voltage distribution for three avalanche
sizes according to equation \ref{eq:P_V_S}.}
\label{fig:size_voltage}
\end{figure}

We also know that the average voltage in the avalanche must be equal
to the avalanche size $S$ divided by the avalanche duration $T\sim
S^{\ts}$, so
\begin{align}
\langle V\rangle &= \int_0^\infty V P(V|S) dV \nonumber \\
		 &=  \int_0^\infty f_{\rm power}(V S^{-y}) dV \nonumber \\
		 &\sim S^{y} \sim S^{1-\ts}.
\label{eq:avg_voltage}
\end{align}
From this, we know that $y=1-\ts$, and the probability of a
voltage $V$ occurring in an avalanche of size $S$ is
\begin{equation}
P(V|S) = V^{-1} f_{\rm power}(V S^{\ts-1}).
\label{eq:P_V_S}
\end{equation}
As can be seen in figure \ref{fig:size_voltage}, this scaling form
works very well.

\subsection{The scaling of avalanche energy $E$ with avalanche size
$S$}

The voltage distribution in equation \ref{eq:P_V_S} allows us to
calculate the dependence of avalanche energy on avalanche size.  The
avalanche energy is simply the average squared voltage, $\langle
V^2\rangle$, times the average avalanche duration, $S^{\ts}$.
Using equation \ref{eq:P_V_S}, this is
\begin{align}
E(S) &= S^{\ts} \int_0^\infty V^2 P(V|S) dV \nonumber \\
	&= S^{\ts} \int_0^\infty V f_{\rm power}(V S^{\ts-1})dV
	\nonumber \\
	&\sim S^{2-\ts}.
\label{eq:E_S}
\end{align}
Note that this is the same result we would find if we assumed that
the time dependence had a square profile.

\subsection{The scaling of the time-time correlation function with
	$S$}

With this information, we can calculate the scaling of the time-time
correlation function within avalanches, which is simply related to the
high-frequency part of the power-spectrum,
$\mathcal{P}_{\text{av}}(\omega)$.  The time-time correlation function
is defined as
\begin{equation}
G(\theta) = \int V(t)V(t+\theta)dt.
\label{eq:tt_correlation}
\end{equation}
If we assume that the magnetic field is increased adiabatically and
the avalanches are well separated in time, we can calculate the
time-time correlation function separately for each avalanche and then
add the individual functions together to get the overall time-time
correlation function.\cite{CrossTerms}

\begin{figure}
\centerline{\includegraphics[width=0.65\textwidth]{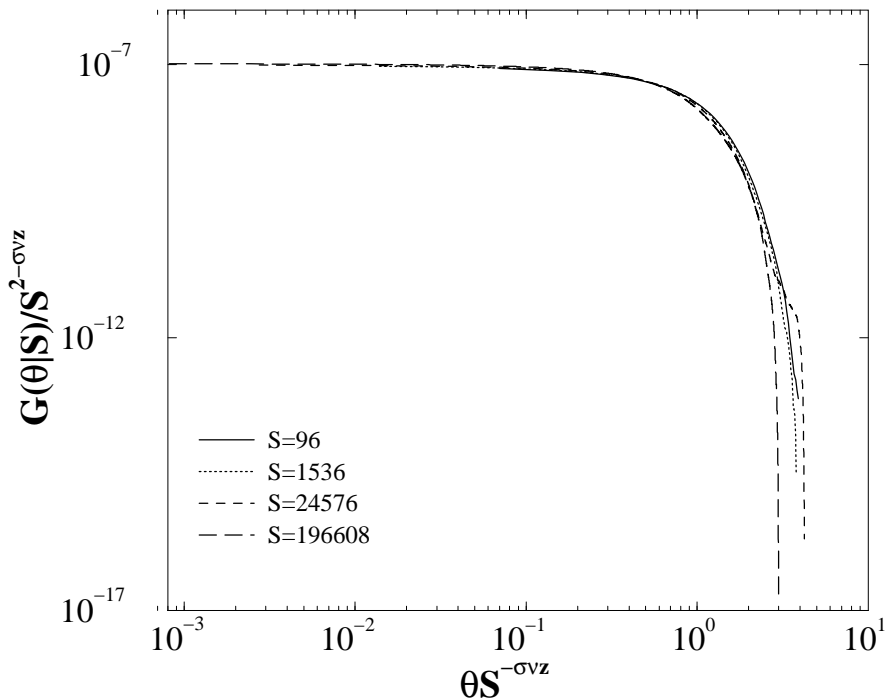}}
\caption{A collapse of avalanche time-time correlation functions
according to equation \ref{eq:tt_correl_S}.  The data is binned
logarithmically to get good statistics.}
\label{fig:time-time}
\end{figure}

This allows us to break up the time-time correlation function into the
contributions from avalanches of different sizes, $S$.  Let
$G(\theta|S)$ be the average time-time correlation function of an
avalanche, \emph{given} that the avalanche is of size $S$.  (In
contrast, the notation $G(\theta,S)$ would denote the contribution of
all avalanches of size $S$ to $G(\theta)$.  This would be weighted by
the probability that an avalanche is of size $S$, $S^{-\tau}$.)  If we
consider equation \ref{eq:tt_correlation} at $\theta=0$, we see that
the $\theta=0$ component of the correlation function is $\int V(t)^2
dt$, which is proportional to the avalanche energy.  Using this fact
along with the scaling of the energy from equation \ref{eq:E_S}, we
find that the time-time correlation function should scale as
\begin{equation}
G(\theta|S) = S^{2-\ts} f_{\rm time\,corr}(\theta S^{-\ts}).
\label{eq:tt_correl_S}
\end{equation}
As shown in figure \ref{fig:time-time}, this scaling works very well over a wide range of avalanche sizes.

\subsection{The scaling of the power spectrum with $S$}

\begin{figure}
\centerline{\includegraphics[width=0.65\textwidth]{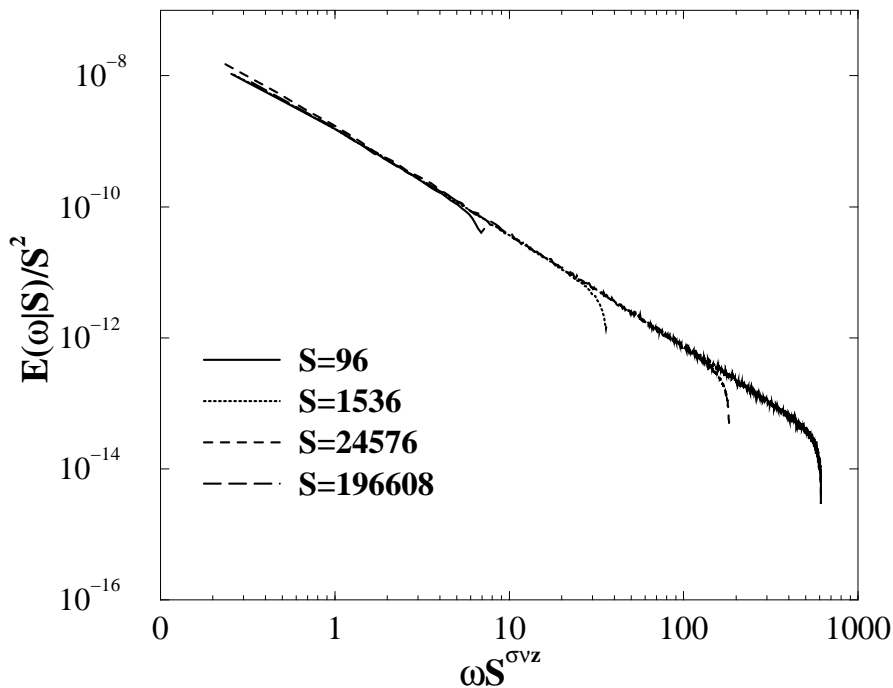}}
\caption{A collapse of the power spectra for different avalanche
sizes according to equation \ref{eq:binned_energy}.}
\label{fig:energy_scale}
\end{figure}

The power or energy spectrum describes the amount of energy released in
Barkhausen noise at each frequency.  It can be calculated as the
cosine transform of the time-time correlation function.
Transforming equation \ref{eq:tt_correl_S}, we find that the scaling
of the power spectrum with avalanche size has the form
\begin{align}
	E(\omega|S) &= \int_0^\infty \cos(\omega\theta) G(\theta|S)
	d\theta \nonumber \\
		&= \int_0^\infty \cos(\omega\theta) S^{2-\ts}
	f_{\rm time\,corr}(\theta S^{-\ts})d\theta \nonumber \\
		&= S^2 f_{\rm time\,corr}(\omega^{1/\ts} S).
\label{eq:binned_energy}
\end{align}
(We'll call this the power spectrum, but use the name $E(\omega)$ to
remind us to divide by the duration of the experiment to change energy
to power.)
A collapse of the power spectra for different avalanche sizes
according to this form is shown in figure \ref{fig:energy_scale}.

\subsection{The power spectrum scaling depends on the large avalanche cutoff}
\label{sec:cutoff_doesnt_scale}

The power spectrum for all avalanches (the quantity usually measured
in experiments) is just the integral $\int D(S) E(\omega|S)dS$, where
$D(S)\sim S^{-\tau}$ is the probability per unit volume of having an avalanche
of size $S$.  If we substitute equation \ref{eq:binned_energy} and 
$D(S)\sim S^{-\tau}$ into the integral, we generate the naive prediction of
\begin{equation}
E_{\rm wrong}(\omega) = \int D(S) E(\omega|S) dS
	\sim\omega^{-(3-\tau)/\ts} \int y^{2-\tau}f_{\rm energy}(y)dy
	\sim\omega^{-(3-\tau)/\ts}
\label{eq:wrong}
\end{equation}
as predicted by Lieneweg {\it et al.},\cite{lieneweg} Dahmen and
Sethna,\cite{sethna2,sethna5,dahmen} and Spasojevi{\'c} {\it et
al.}\cite{stanley2}. In the next section \ref{sec:linear_in_S} we will
argue that the energy contributed by an avalanche at a fixed frequency
$\omega$ is proportional to $S$: this immediately implies using the scaling
form \ref{eq:binned_energy} that $f_{\rm energy}(y) \sim A/y$ for large $y$.
Thus the integrand in equation \ref{eq:wrong} dies off as a power law
$y^{1-\tau}$ for large $y$. Hence, it gives the correct answer if
$\tau>2$, but for $\tau<2$ the naive indefinite integral in
\ref{eq:wrong} diverges and must be replaced by a definite integral. The
naive result should work for the integrated avalanche size distribution
for the short-range model\cite{sethna2,sethna4} at the critical
point $R_c$. In that case (which we won't study in this paper), the
corresponding exponent $\tau + \sigma \beta \delta = 2.03$ in three
dimensions (close to two, so we'd expect logarithmic corrections), and
rising to the mean-field value of $9/4$ in six dimensions. For the
models we study here, $\tau<2$ and the cutoff at the largest avalanches
in the integral over avalanche sizes changes the form of the power
spectrum.  In the section \ref{sec:int_energy} after next, we will
derive our resulting exponent relation for the power spectrum.

%

\subsection{The energy at frequency $\omega$ scales linearly
with the avalanche size $S$.}
\label{sec:linear_in_S}
 
\begin{figure}
\centerline{\includegraphics[width=0.65\textwidth]{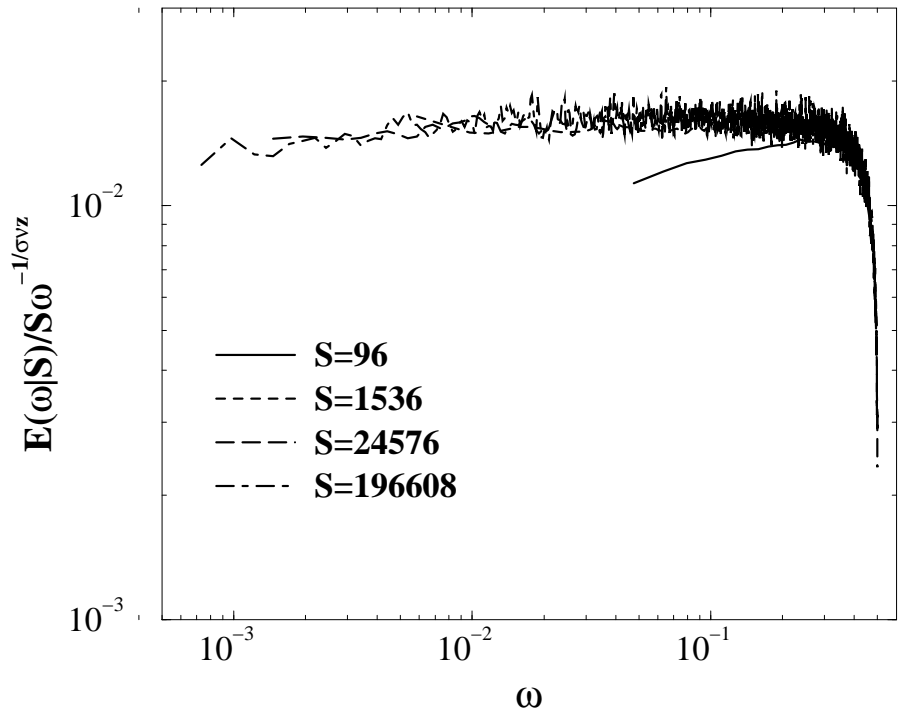}}
\caption{A collapse of the power spectra for several avalanche sizes
$S$.  The collapse is performed by dividing out the linear $S$
dependence.  The curves are made flat by dividing out the simple power
law $\omega^{1/\ts}$.  The $\omega$ axis has not been
rescaled, so the high $\omega$ cutoffs collapse together and the small
$\omega$ cutoffs do not.}
\label{fig:energy_scale2}
\end{figure}

We shall argue in this section that except at very small frequencies the
energy at a fixed frequency $\omega$ is proportional to $S$.  According
to equation \ref{eq:binned_energy}, this implies that the power spectrum
at fixed $S$ scales as $\omega^{-1/\ts}$.  This means that dividing
$E(\omega|S)$ by $S\omega^{-1/\ts}$ should collapse the curves and
eliminate the $\omega$ dependence. The result of this collapse for the
infinite-range model can be seen in figure \ref{fig:energy_scale2}.

The fact that the energy at frequency $\omega$ scales linearly with
$S$ follows from two assumptions.  First, each spin in the avalanche
contributes equally to $E(\omega)$.  Second, the contribution of a
given spin is independent of the size of the avalanche it is in.  (Of
course, a few spins at the beginning of the avalanche might differ, so
long as the fraction of such spins vanishes for large avalanches.)
One way in which this could be true would be if the following two
hypotheses were true.  First, each spin contributes to $E(\omega)$
only through its correlations with physically nearby spins.  Second,
the local growth of an avalanche does not reflect the overall size of
the avalanche.  The first hypothesis seems likely in the models
described in this paper because the times of physically distant spin
flips are likely to be randomly distributed in time and contribute
incoherently to the power spectrum.  Only nearby spin flips will be
correlated in time and contribute coherently.  The second hypothesis
is also likely to be true because the avalanches are occurring at a
critical point.  Every part of the avalanche is always on the verge of
stopping regardless of the size of the avalanche.

We can check these hypotheses by breaking up the power spectrum into
the contributions from pairs of spins at different radii $r$.  The
time-time correlation function for avalanches of size $S$ can be
written in terms of individual pairs of spins as
\begin{equation}
	G(\theta|S) = \sum_{i,j} \delta(t_j-t_i-\theta),
\end{equation}
where $t_i$ is the time at which spin $i$ flips.  From this form, we can
use an additional delta function to pull out the contribution due to pairs of spins separated by a
distance $r$:
\begin{equation}
	G(\theta,r|S) = \left\langle \sum_j \delta(t_j-t_i-\theta)
	\delta(|\vec{r}_j-\vec{r}_i|-r)\right\rangle_i,
\end{equation}
where $\langle\rangle_i$ implies an averaging over all values of $i$.
Using this definition of the function $G(\theta,r|S)$, we can rewrite
the time-time correlation function as
\begin{equation}
	G(\theta|S) = S \int G(\theta,r|S) dr.
	\label{eq:r_correl}
\end{equation}

Now, we can use equation \ref{eq:r_correl} to calculate the
contribution of spins separated by a distance $r$ to the power
spectrum $E(\omega|S)$ at a frequency $\omega$.  Taking the cosine
transform of equation \ref{eq:r_correl}, we find that
\begin{align}
	E(\omega|S) &= S\int\int\cos(\omega\theta) G(\theta,r\,|S)dr
	d\theta \nonumber \\
		&\equiv S\int E(\omega,r\,|S) dr,
\label{eq:e_w_r_s}
\end{align}
where the function $E(\omega,r\,|S)$ is defined by this equation.
Notice that $E(\omega,r\,|S)$ must have a cutoff at the largest $r$
present in an avalanche of size $S$.  We can see from equation
\ref{eq:e_w_r_s} that in order for $E(\omega|S)$ to be proportional to
$S$, the integral must be independent of this $S$ dependent cutoff.
This is a more precise statement of the hypothesis that only
correlations between nearby spins contribute to the power spectrum.
It is also necessary that except for extreme values of $\omega$ and
$r$, $E(\omega,r\,|S)$ must be independent of $S$.  Combined with the
first condition, this corresponds to the hypothesis that the local
growth of the avalanche should not reflect the overall size of the
avalanche.

\begin{figure}
\centerline{\includegraphics[width=0.65\textwidth]{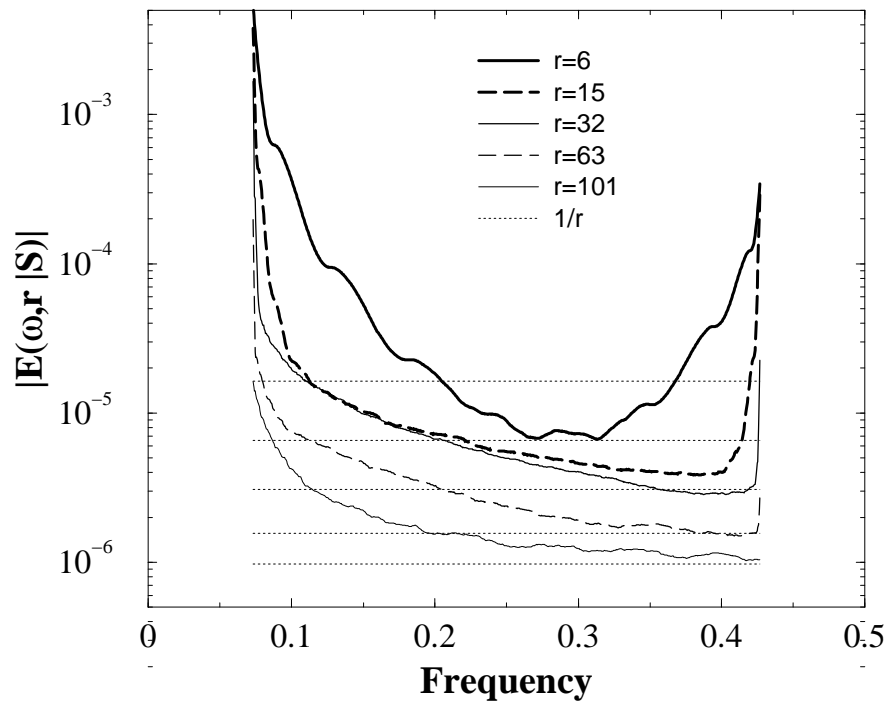}}
\caption{This graph shows the function $|E(\omega,r\,|S)|$ for a range
of values of $r$ at $S=32500$.  The function $E(\omega,r\,|S)$ not only
decays with $\omega$, but also oscillates about zero.  To better
compare the amplitudes of the curves at different $r$, we took the
absolute value and performed a running average over $\omega$,
averaging out the oscillations.  The horizontal lines show how the
positions of these curves should scale if they went as $1/r$.  Notice
that while for small $r$ the amplitudes drop slower than $1/r$, for
larger $r$ the amplitudes drop off approximately as $1/r$.}
\label{fig:e_w_r_s}
\end{figure}

\begin{figure}
\centerline{\includegraphics[width=0.65\textwidth]{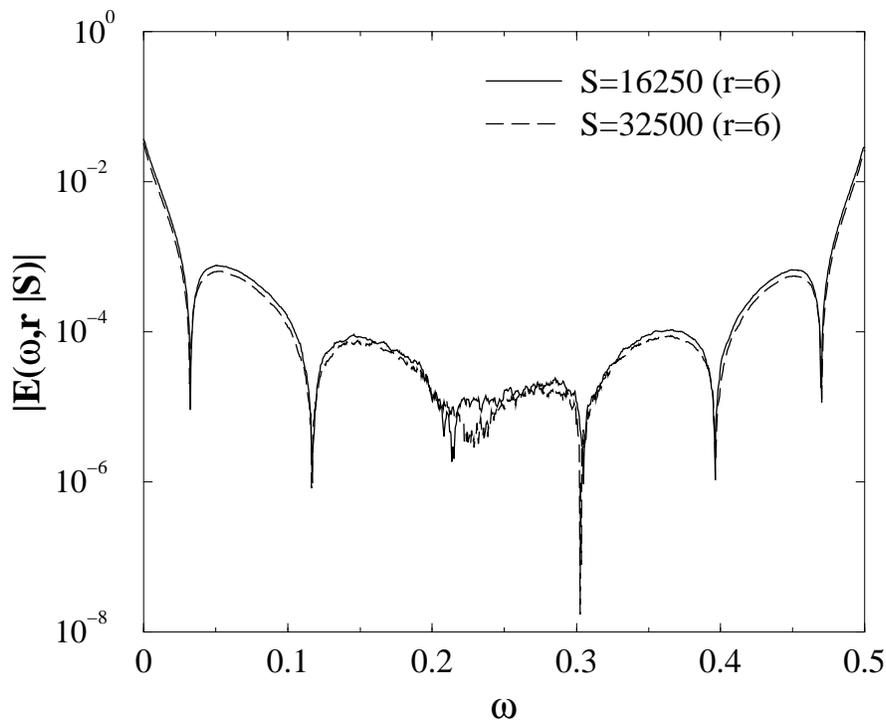}}
\caption{This graph shows $|E(\omega,r\,|S)|$ at $r=6$ for two sizes
of avalanches.  No averaging has been performed, so the oscillations
are visible (the function changes sign at each dip).  For larger $r$,
the oscillations are much faster.
Notice that the curves are nearly identical for the two
sizes---$E(\omega,r\,|S)$ has no significant $S$ dependence at small
$r$.}
\label{fig:e_w_r_s2}
\end{figure}

Figure \ref{fig:e_w_r_s} shows $E(\omega, r\,|S)$ is
decaying approximately as $1/r$.  Figure \ref{fig:e_w_r_s2} shows that
it is independent of $S$ and that the decay is oscillating about
zero.  Neighboring radial shells contribute with opposite sign.
Hence the integral \ref{eq:e_w_r_s} appears to be conditionally
convergent at large distances: each spin contributes to $E(\omega)$
only through its correlations with nearby spins, and $E(\omega|S) \sim
S$.
\subsection{Integrating the power spectrum over avalanche sizes}
\label{sec:int_energy}

In section \ref{sec:cutoff_doesnt_scale} we found that we need to
understand the cutoff at large avalanche sizes. In our simulations
there is a typical largest avalanche size $S^*$ which is primarily due to
finite size effects: $S^* \sim L^{1/{\sigma\nu}}$.  In experiments, the
typical largest avalanche size could be determined by the finite experimental
duration or from the demagnetizing forces.  (In some cases, demagnetizing
forces also contribute to the cutoff in our simulations.) These 
effects would cut off the probability of getting large avalanches.

The contribution to $E(\omega)$ from an avalanche of size $S$ scales linearly
in $S$, for large avalanches: each spin contributes the same amount. The
cutoff at large avalanche sizes dominates the scaling of $E(\omega)$ precisely
because most spins are in the largest avalanches.
The distribution of avalanche sizes $D(S,S^*)$ cannot be a simple power law
cut off at $S^*$ for $\tau<2$, because we know $\int S D(S,S^*) dS = 1$;
$\int^{S^*} S^{1-\tau} dS$ depends on $S^*$ and diverges as $S^*\to\infty$
for $\tau<2$. Hence the overall amplitude of $D$ must decreases as $S^*$ gets
bigger.  If we choose a scaling form for the avalanche size distribution with 
an overall amplitude which depends upon $S^*$, we can fix this problem:
\begin{align}
	D(S,S^*) dS &= (S^*)^{-2} f_{\rm size}(S/S^*) \\
	f_{\rm size}(y) &\sim y^{-\tau}	~~~y \to 0 \\  
	\int S D(S,S^*) dS &= \int (S^*)^{-2} S f_{\rm size}(S/S^*) dS
			   = \int y f_{\rm size}(y) dy = 1 
	\label{eq:size_normalization}
\end{align}
where the last equation provides a condition on the scaling function
$f{\rm size}$.

We are now ready to integrate the power spectrum over $S$, finding the
form $E(\omega)$ corresponding to equation \ref{eq:wrong} but valid for
$\tau<2$. We discovered in the last sections that $f_{\rm energy}(y) \to A/y$.
We control the integral by adding and subtracting $A/y$: 
\begin{align}
 E(\omega) &= \int D(S,S^*) E(\omega|S) dS \nonumber \\
	   &= \int [(S^*)^{-2} f_{\rm size}(S/S^*)] 
			[S^2 f_{\rm energy}(\omega^{1/\ts} S)] dS \nonumber \\
	   &= \int (S/S^*)^2 f_{\rm size}(S/S^*) 
			 f_{\rm energy}(\omega^{1/\ts} S) dS \nonumber \\
	   &= \int (S/S^*)^2 f_{\rm size}(S/S^*) 
			 A/(\omega^{1/\ts} S) dS \nonumber \\
	   &+ \int (S/S^*)^2 f_{\rm size}(S/S^*) 
			 [f_{\rm energy}(\omega^{1/\ts} S) 
				     - A/(\omega^{1/\ts} S)] dS 
\label{eq:energy_spectrum1}
\end{align}
In the first integral, we set $z=S/S^*$. In the second integral, we
set $y=\omega^{1/\ts} S$. Also, the second integral now converges, so
we can substitute $f_{\rm size}(S/S^*) \to (S/S^*)^{-\tau}$:
\begin{align}
 E(\omega) &= \omega^{-1/\ts} \int A z^2 f_{\rm size}(z) dz
		+ (S^*)^{\tau-2} \omega^{-(3-\tau)/\ts} 
			\int y^{2-\tau} [f_{\rm energy}(y)-A/y] dy 
\label{eq:energy_spectrum2a} \\
	   &\sim \omega^{-1/\ts}	~~~~~~ (\tau < 2) 
\label{eq:energy_spectrum2b}
\end{align}

\begin{figure}
\centerline{\includegraphics[width=0.65\textwidth]{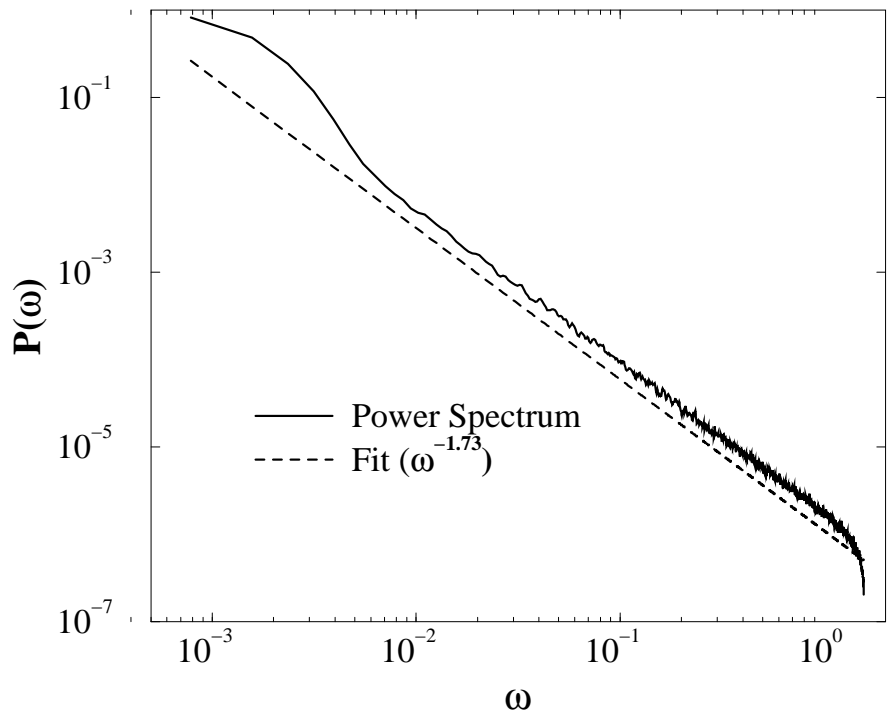}}
\caption{The power spectrum for the infinite range model.  The
dashed line is not a fit.  It is a power law with an exponent
$1/\ts$.}
\label{fig:power_scaling}
\end{figure}

The exponent in the second
term of equation \ref{eq:energy_spectrum2a}, $(3-\tau)/\ts$, is 
$E_{\rm wrong}$ of
equation \ref{eq:wrong}, given by ignoring the cutoff at large avalanche
sizes.  However, for $\tau<2$ the first term will
dominate over the second term both for large system
sizes and for large $\omega$.  Only for $\tau>2$ will the second term
dominate. For all of the models we are considering, $\tau<2$.  (See
table \ref{tab:exponents}.) In fact, the exponent $(3-\tau)/\ts$
disagrees badly with the results observed in both simulations and
experiments, and the exponent $1/\ts$ agrees very well.  In mean field
theory, we can actually derive rigorously that the power spectrum
exponent is $2$ (see appendix \ref{app:mf}).  This is in perfect
agreement with the exponent $1/\ts$, and complete disagreement with the
exponent $(3-\tau)/\ts$, which would be 3.  A plot of the power law for
the infinite-range model can be seen in figure \ref{fig:power_scaling}.
A comparison of the two possible exponents with simulations for each of
the three models can be seen in table \ref{tab:comparison}.

\begin{table}
\begin{center}
\begin{tabular}{lccc}
\hline\hline
	& $1/\ts$ & $(3-\tau)/\ts$ & Simulated \\
\hline
Short-range	& 1.72	& 2.41	& 1.70 \\
Infinite-range	& 1.72	& 3.00	& 1.70 \\
Dipole		& 2.00	& 3.00	& 2.00 \\
\hline\hline
\end{tabular}
\end{center}
\caption{This table compares values of the power spectrum exponent
measured in simulations with the values predicted by the exponents
$1/\ts$ and $(3-\tau)/\ts$.  Notice that for all three
models $1/\ts$ is very close to the simulated exponent, and
$(3-\tau)/\ts$ is in complete disagreement.}
\label{tab:comparison}
\end{table}

\section{How universal is the exponent $1/\sigma \nu \lowercase{z}$?}
\label{sec:snz}

\begin{figure}
\centerline{\includegraphics[width=0.65\textwidth]{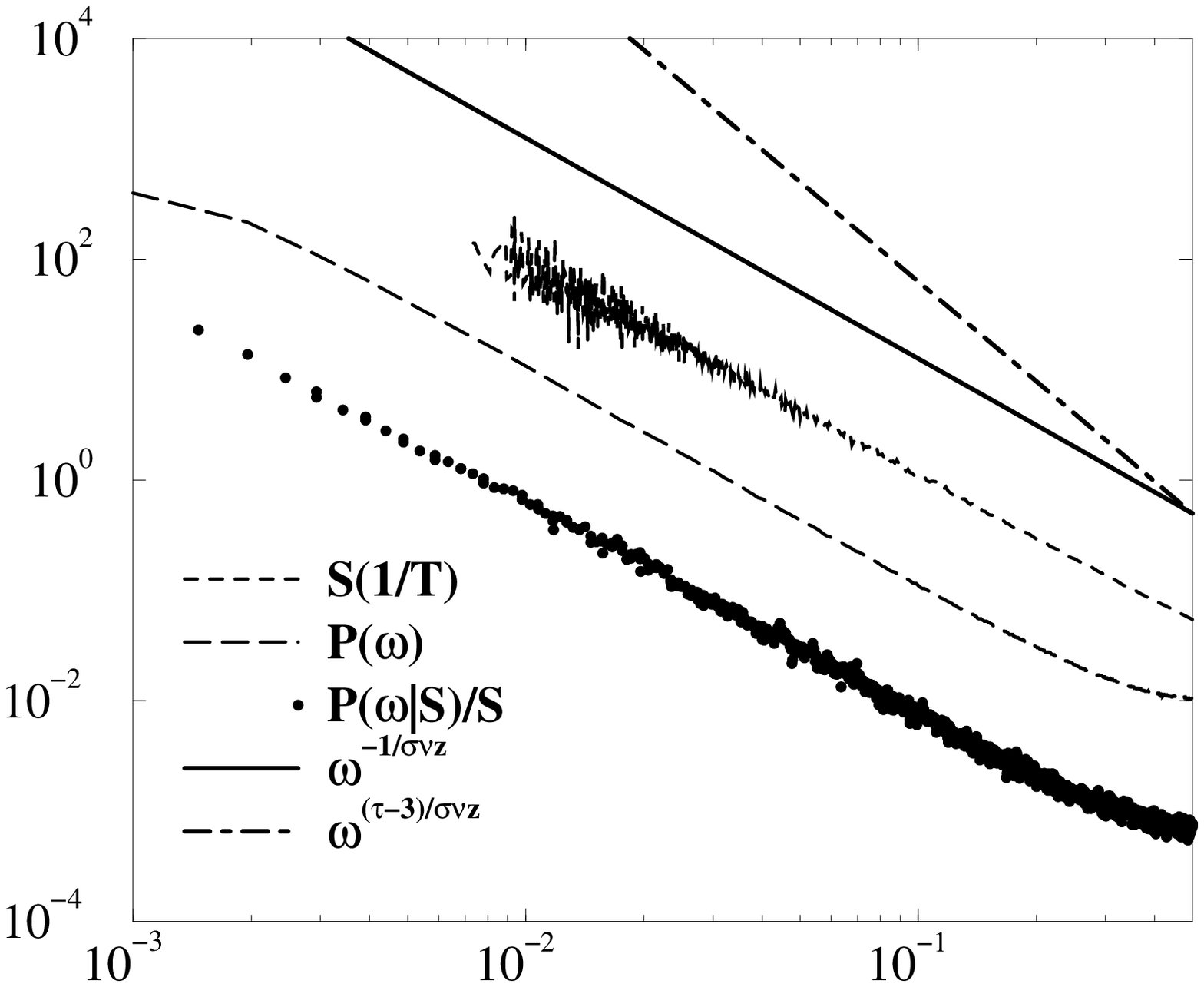}}
\caption{The mean field model.  Notice that the power spectrum for
avalanches of size $S$ collapses when divided by $S$, and that the power
spectrum for avalanches of size $S$ and the overall power spectrum
both have the same power law as the average avalanche size as a
function of the inverse avalanche duration. Sizes 96, 1536, 24576, and
196608 are shown in the collapse of $P(\omega|S)/S$. The average avalanche
size as a function of the inverse avalanche duration is the definition
of the exponent $1/\ts$.  Power laws of $1/\ts$ and
$(3-\tau)/\ts$ are also shown for comparison.  Notice that
$(3-\tau)/\ts$ is completely incompatible with the results.}
\label{fig:mf_scale}
\end{figure}
\begin{figure}
\centerline{\includegraphics[width=0.65\textwidth]{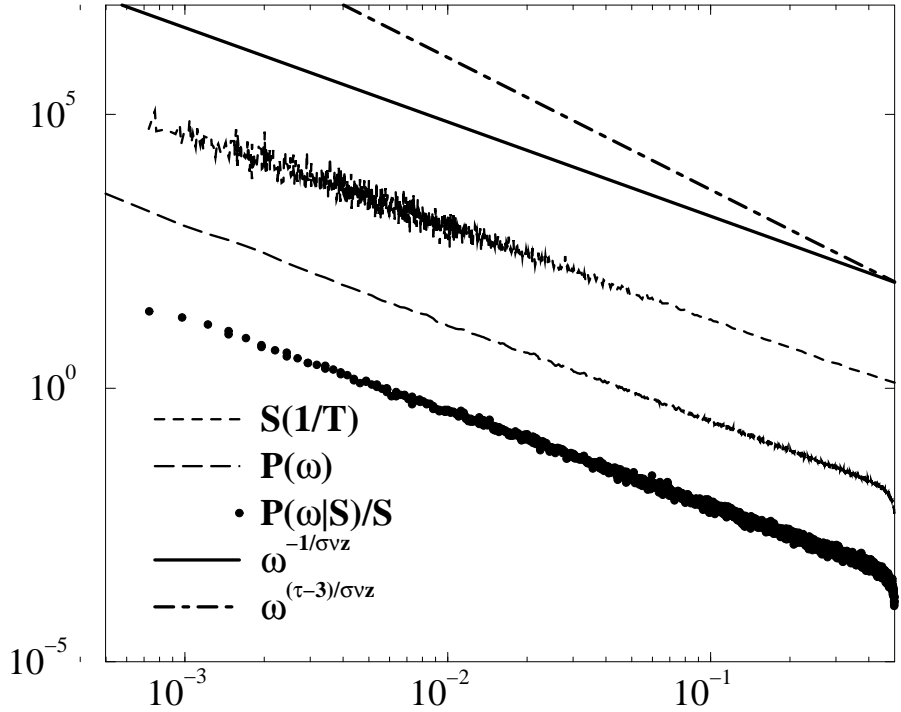}}
\caption{The short-range model in 3 dimensions.  The plots and the sizes
are the same as in figure \ref{fig:mf_scale}.}
\label{fig:3Dshort_scale}
\end{figure}

So far in this paper, we have only discussed these results for the infinite
range model in three dimensions.  However, we have also checked these
results in several other situations where both $\tau$ and $\ts$ take on
a range of values.  (In all cases, $\tau<2$.)  We have checked the
short-range model in 3 and 4 dimensions, the infinite-range model in 3
and 4 dimensions, and mean field theory.  In all cases, the linear
scaling with $S$ seems exact, and both $E(\omega|S)$ and $E(\omega)$
scale as $\omega^{-1/\ts}$ to within simulational precision.  In all
cases, the exponent $(3-\tau)/\ts$ is completely inconsistent with the
observed results.  The results of mean field theory and the
short-range model at the three dimensional critical point can be seen
in figures \ref{fig:mf_scale} and \ref{fig:3Dshort_scale}.  The
results for the four dimensional short-range model and the four
dimensional infinite range model were also completely consistent with
a $1/\ts$ scaling exponent.

\section{Discussion}
\label{ap:prev}
There have been several previous predictions of the power spectrum
exponent.  In her Ph.D. thesis, Karin Dahmen\cite{dahmen} did a
calculation similar to the one done in this paper, but she ignored the
problems with the integral and came up with the exponent
$(3-\tau)/\ts$, which we have shown should hold only if $\tau>2$.
This exponent was published by our group\cite{sethna2} without the derivation,
and was compared to experimental work by Cote and Meisel\cite{cote1,cote2}
and Bertotti {\it et al.}\cite{bertotti} 
The correct exponent form derived here makes the
agreement between the short-range model and these experiments
substantially better: they quote an exponent of ``around 2'', our
former (wrong) prediction was $2.46\pm 0.17$, and the correct
prediction is $1.75\pm 0.25$.  

Presumably the earliest derivation of
$(3-\tau)/\ts$ (we thank the referee for pointing this out), is by
Lieneweg and Gross-Nobis\cite{lieneweg} in 1972: their exponent 
$1+2\epsilon -\delta$ translates precisely to $(3-\tau)/\ts$ in
our notation.\cite{lieneweg} They use this to analyze their unannealed
data, for which they measure $\tau = 2.1$ (hence making their derived exponent
appropriate); their annealed wires show values of $\tau = 1.73$, and indeed
their power spectrum shows a much steeper falloff, quite close to our
predicted value $1/\ts$ which they measure to be $1.63$.

Spasojevi{\'c}, Bukvi{\'c},
Milosevi{\'c} and Stanley\cite{stanley2} also came up with the same
(wrong) form based on arguments about the average pulse shape.  
They measured an experimental power spectrum exponent of $1.6$. For
their experimentally determined values of $\tau=1.77$ and $\ts=0.662$,
they found that $(3-\tau)/\ts=1.86$ fit their data well, but the
value of $(3-\tau)/\ts=2.46$ quoted by Dahmen et. al.\cite{sethna2}
was much too large.  This lead them them to disregard the plain old
critical model with domain nucleation as a possible explanation of
Barkhausen noise in their system.  However, note that their
experimental value of $1/\ts=1.51$ is as close to their measured power
spectrum exponent as their value of $(3-\tau)/\ts=1.86$.  The value of
$1/\ts=1.75\pm 0.25$ quoted by Dahmen {\it et al.} is even closer to
the experimentally observed power spectrum exponent.  (The value of
$\tau=1.6$ predicted by Dahmen {\it et al.} is also closer to the
experimentally measured $\tau=1.77$ than any of the other models:
front propagation has an exponent $\tau=1.3$, and mean-field theory
has an exponent $\tau=1.5$.)  One should note, however, that the
average avalanche shapes measured by Spasojevi{\'c} {\it et al.}
disagree with those predicted by the short-range model (figure
\ref{fig:avg_aval}).

In 1989, Jensen, Christensen and Fogedby\cite{jcf} published a
different calculation of the power spectrum exponent for sandpile
models, which has been cited many times as an explanation for
$1/\omega^2$ scaling of the power spectrum in sand-piles, Barkhausen
noise, and other systems.  They made two major assumptions in their
derivation.  First, they assumed that the avalanche shape could be
approximated by a box function: $V(t)=S/T$ for all $t<T$.  For our
models, this assumption turns out to be valid for calculating the
average avalanche energy, but is not valid for determining the overall
scaling of the time-time correlation function.  Second, they assumed
that one of their scaling functions, which was related to the
time-time correlation function, had the simple scaling form $G(T)\sim
T^\alpha \exp(-T/T_0)$.  As can be seen from the exact mean field
time-time correlation function in equation \ref{eq:mftime_time_final},
this is not at all a safe assumption.  In terms of our exponents
$\tau$ and $\ts$, their prediction was that the power spectrum would
have the form $E(\omega)\sim \omega^{(3-\tau)/\ts}$ for $\tau>3-2\ts$
and $E(\omega)\sim\omega^{-2}$ for $\tau<3-2\ts$.  This is the same as
our result in the case where $\ts=1/2$.  For the particular model they
were studying, their assumptions seem to be correct, but in the more
general case, the exponent $(3-\tau)/\ts$ should become invalid for
$\tau<2$, rather than $\tau<3-\ts$.  Also, for $\tau<2$, exponents
other than $2$ are possible, depending on the particular dynamics.
For the class of models described in this paper, the correct exponent
for $\tau<2$ is $1/\ts$.

Narayan\cite{narayan2} derives a power-spectrum exponent for front
propagation, considering the dynamic height-height correlation function
for the growing interface. His exponent\cite{NarayanPC}, $(2 \zeta +
d-1)/z -1$ involves the static height-height correlation exponent
$\zeta$: $\langle h(x,t) h'(x',t)\rangle \sim |x-x'|^{2 \zeta}$, along
with the dimension $d$. This prediction is rather different from the two
discussed in this paper. Narayan and Fisher's $5-d=\epsilon$-expansion
results\cite{narayan1} show $\zeta=2/3$ and $z = 2 - 2 \epsilon/9
+O(\epsilon^2)$ yielding a predicted power-spectrum exponent of $8/7 +
O(\epsilon^2)$ in three dimensions, quite different from the value of
around 1.7 that we measure. Narayan's derivation does not separate the
high-frequency power law $\mathcal{P}_{\text{av}}(\omega)$ which we
study from the low-frequency power law
$\mathcal{P}_{\text{corr}}(\omega)$ reflecting the correlations between
avalanches (discussed in the introduction): his analysis is probably
describing the asymptotic behavior of the latter.

There are other specific models for which the power spectrum exponent
has been calculated.  For example, Bak, Tang and
Wiesenfeld\cite{sand_piles} calculated the power spectrum exponent for
their sandpile models, and Durin, Bertotti, and Magni\cite{Durin}
calculate the form and asymptotic power law for the power-spectrum of the
ABBM\cite{ABBM1,ABBM2} model.  Without further investigation, we can't expect
our arguments for an exponent of $1/\ts$ to hold for these and other
models.  However we do expect that in any avalanche based model, the
power spectrum exponent for $\tau>2$ will be $(3-\tau)/\ts$, and
another exponent will dominate for $\tau<2$.  Whenever the arguments
in section \ref{sec:linear_in_S} hold, for $\tau<2$ the
exponent $1/\ts$, relating the avalanche size to avalanche duration, should
dominate.

\acknowledgments
We acknowledge useful conversations and correspondence with
Karin Dahmen, and helpful comments from Chris Henley. We thank Onnutom Narayan
for a careful reading of the manuscript, and for pointing out a confusion
in our argument.  This work was supported by NSF DMR \#9805422.

\appendix
	\section{The mean field power spectrum}
	\label{app:mf}
	In mean field theory, we can calculate the power spectrum exactly.
	The Hamiltonian in mean field theory is the Hamiltonian in equation
	\ref{eq:hamiltonian}, without nearest neighbor or dipole terms:
	\begin{equation}
	\label{eq:mfhamiltonian}
	\mathcal{H} = - \sum_i (H+J+h_i)s_i.
	\end{equation}
	When a spin flips with an external field of $H$, all spins with random
	fields between $-(H+JM)$ and $-(H+J(M+2/N))$ will flip.  Therefore,
	each spin has a probability of $\frac{2J}{N} \rho(-H-JM)$ of flipping,
	where $\rho(h)$ is the probability distribution of the random fields.
	On average, $2J\rho(-H-JM)$ spins will be flipped.  If
	$2J\rho(-H-JM)>1$, then the avalanche will tend to grow indefinitely,
	and there will be an infinite avalanche.  If $2J\rho(-H-JM)<1$, then
	the avalanche will quickly die out, and all avalanches will be small.
	If $2J\rho(-H-JM)=1$, then avalanches will always be finely balanced
	between continuing and dying, and there will be a critical
	distribution of avalanches.  If the random field distribution
	$\rho(h)$ has a maximum value of $1/2J$, then there will be a critical
	distribution of avalanches at the value of $H=H_c$ where $\rho(h)$ is
	a maximum.

	Let us calculate the power spectrum for the critical system with
	$\rho(h)=1/2J$.  To begin with, we will calculate the probability
	distribution of time-series ${n_1,n_2,\dots,n_\infty}$.  We know that
	in shell zero, exactly $n_0=1$ spins will flip.  At the critical
	point, where each spin on average causes one more spin to flip, shell
	one will have a Poisson distribution with mean one: $P(n_1)=\frac{1}{e
	n_1!}$.  Shell $i$ will have a Poisson distribution with mean
	$n_{i-1}$: $P(n_i) = \frac{e^{n_i-1} n_{i-1}^{n_i}}{n_i!}$.  Therefore,
	the probability distribution for the entire time series will be
	\begin{equation}
	P(1,n_1,n_2,\dots,n_\infty) = \frac{1}{e n_1!} \prod_{i=2}^\infty
	\frac{e^{n_{i-1}} n_{i-1}^{n_i}}{n_i!}.
	\label{eq:mftime_series}
	\end{equation}

	Now, from equation \ref{eq:mftime_series}, we can calculate the average
	time-time correlation function
	\begin{equation}
	G(\theta) = \sum_{i=0}^\infty \langle n_i n_{i+\theta} \rangle,
	\label{eq:mftime_time}
	\end{equation}
	where
	\begin{equation}
	\langle n_i n_{i+\theta} \rangle = \sum_{\{n1,\dots,n_\infty\}=0} n_i
	n_{i+\theta} P(n_1,\dots,n_\infty).
	\label{eq:mf_correl}
	\end{equation}
	To simplify equation \ref{eq:mf_correl}, we need to use several
	properties of the Poisson distribution: $\sum_{n=0}^\infty
	\frac{e^{-x} x^n}{n!} = 1$, $\sum_{n=0}^\infty n \frac{e^{-x} x^n}{n!}
	= x$, and $\sum_{n=0}^\infty n^2 \frac{e^{-x} x^n}{n!} = x+x^2$.
	Using the first rule repeatedly, we can simplify equation \ref{eq:mf_correl} to
	\begin{equation*}
		\langle n_i n_{i+\theta} \rangle =
		\sum_{\{n_1,\dots,n_{i+\theta}\}=0}^\infty n_i n_{i+\theta} P(n_1,\dots,n_{i+\theta}).
	\end{equation*}
	Then, using the second rule repeatedly, we can further simplify to
	\begin{equation*}
		\langle n_i n_{i+\theta} \rangle =
		\sum_{\{n_1,\dots,n_i\}=0}^\infty n_i^2 P(n_1,\dots,n_i).
	\end{equation*}
	Now, applying the second and third rules repeatedly, we can simplify
	to a single sum:
	\begin{align}
		\langle n_i n_{i+\theta} \rangle &= \sum_{n_1=0}^\infty
		\frac{(i-1) n_1 + n_1^2}{e n_1!} \nonumber \\
			&= i+1.
	\label{eq:mf_correl_simp}
	\end{align}
	Notice that the correlation between two times is proportional only to
	the first time, and not the separation between the times.

\begin{figure}
	\centerline{\includegraphics[width=0.65\textwidth]{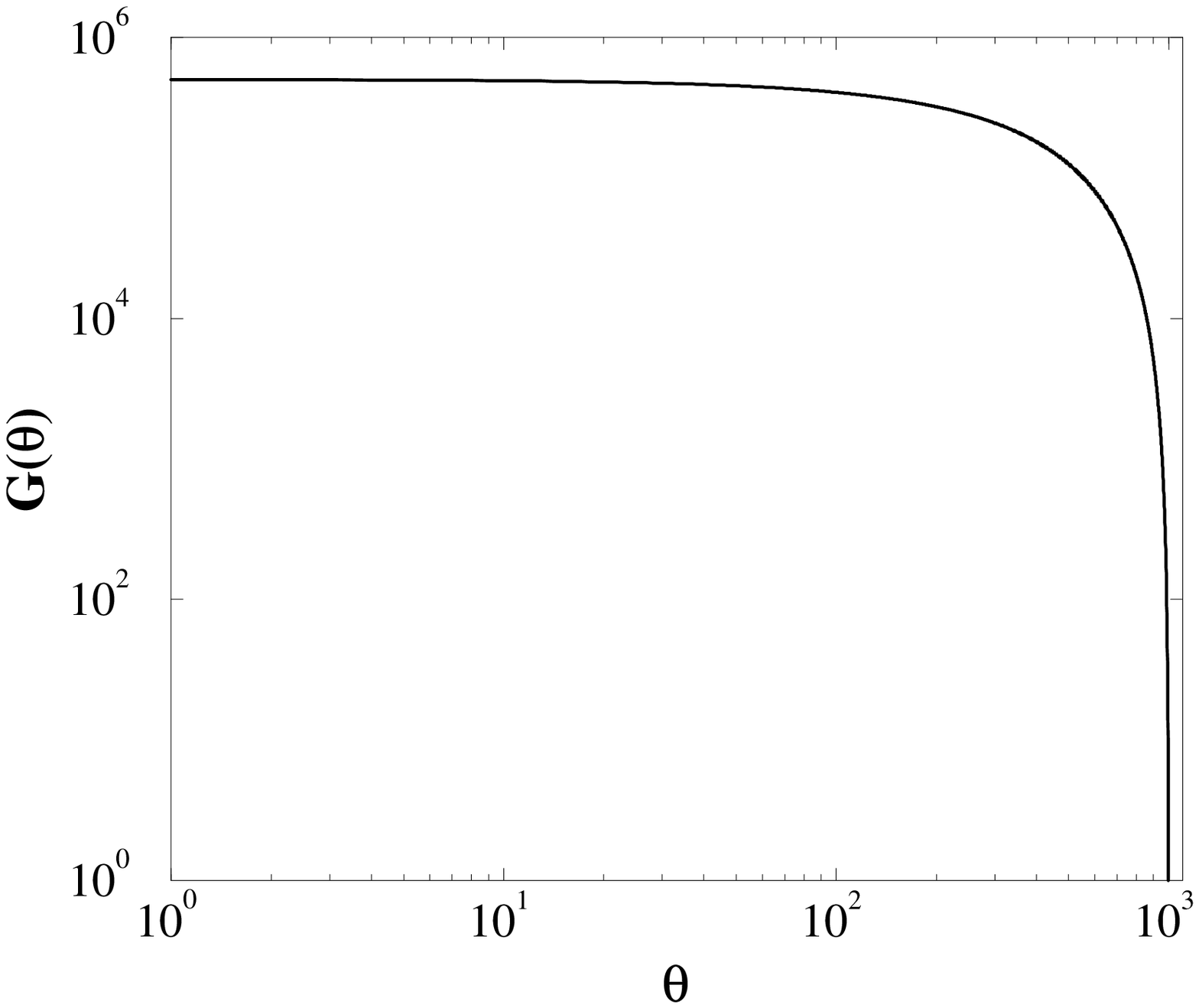}}
	\caption{The exact time-time correlation for mean field theory with a
	maximum time cutoff of $T=1000$.}
	\label{fig:mftime-time}
	\end{figure}

	Now, we can find the value of the time-time correlation function
	$G(\theta)$ by summing the result of equation \ref{eq:mf_correl_simp}.
	Because the time-time correlation function as defined is proportional
	to the square of the total time, we must cut off the summation at some
	maximum time $T$ to get a finite result.  Summing according to
	equation \ref{eq:mftime_time}, we find
	\begin{align}
	G(\theta) &= \sum_{i=0}^{T-\theta-1} i+1 \nonumber \\
		&= \frac{(T-\theta)(T-\theta+1)}{2} \nonumber \\
		&=
		\frac{T^2}{2}+\frac{T}{2}+\frac{\theta^2}{2}-\frac{\theta}{2}-T\theta.
	\label{eq:mftime_time_final}
	\end{align}
	This shape for a cutoff time of $T=1000$ is shown in figure
	\ref{fig:mftime-time}.  Notice that the exact shape in mean-field
	theory is very similar to the experimentally measured correlation
	functions in three dimensions.

	To get the exact form of the mean-field power spectrum, we just take
	the cosine transform of equation \ref{eq:mftime_time_final}.
	\begin{align}
		E(\omega) &= \int_0^\infty \cos(\omega\theta)G(\theta)d\theta
		\nonumber \\
			&= \int_0^\infty \cos(\omega\theta)\left(\frac{T^2}{2}
		+ \frac{T}{2} + \frac{\theta^2}{2} - \frac{\theta}{2} -
		T\theta\right)d\theta \nonumber \\
			&= T \omega^{-2} + \frac{\omega^{-2}}{2}
		(1-\cos(T\omega))-\omega^{-3}\sin(T\omega).
	\label{eq:mf_energy}
	\end{align}
	The dominant term in this equation is $T\omega^{-2}$.  This is the
	same exponent as predicted by the general scaling arguments:
	$\frac{1}{\ts}=2$.  The general scaling arguments also
	predicted that a term smaller by a factor of $T$ and with the exponent
	$(3-\tau)/\ts=3$ would be subtracted off, and it is, but
	multiplied by a factor of $\sin(T\omega)$.  However, the
	$\sin(T\omega)$ turns out to simplify things even more.  Because the
	correlation function was actually discrete, we should consider a
	discrete power spectrum, where the frequencies are multiples of
	$\omega_0=\frac{2\pi}{T}$.  This means that $\cos(T\omega)=1$ and
	$\sin(T\omega)=0$ for all $\omega$ in the discrete spectrum.  Because
	of this, all terms except the $T\omega^{-2}$ term drop out.  If we
	divide by $T$ to get the power spectrum, we find
	\begin{equation}
		\mathcal{P}(\omega) = \omega^{-2}.
	\label{eq:mfpower}
	\end{equation}

	\bibliography{power_spectrum}

\begin{thebibliography}{10}
\expandafter\ifx\csname bibnamefont\endcsname\relax
  \def\bibnamefont#1{#1}\fi
\expandafter\ifx\csname bibfnamefont\endcsname\relax
  \def\bibfnamefont#1{#1}\fi
\expandafter\ifx\csname url\endcsname\relax
  \def\url#1{\texttt{#1}}\fi
\expandafter\ifx\csname urlprefix\endcsname\relax\def\urlprefix{URL }\fi
\providecommand{\bibinfo}[2]{#2}
\providecommand{\eprint}[2][]{\url{#2}}

\bibitem{sc}
\bibinfo{author}{\bibfnamefont{S.}~\bibnamefont{Field}},
  \bibinfo{author}{\bibfnamefont{J.}~\bibnamefont{Witt}},
  \bibinfo{author}{\bibfnamefont{F.}~\bibnamefont{Nori}}, \bibnamefont{and}
  \bibinfo{author}{\bibfnamefont{X.}~\bibnamefont{Ling}},
  \bibinfo{journal}{Phys. Rev. Lett.}
  \textbf{\bibinfo{volume}{74}}(\bibinfo{number}{7}), \bibinfo{pages}{1206}
  (\bibinfo{year}{1995}).

\bibitem{sand_piles}
\bibinfo{author}{\bibfnamefont{P.}~\bibnamefont{Bak}},
  \bibinfo{author}{\bibfnamefont{C.}~\bibnamefont{Tang}}, \bibnamefont{and}
  \bibinfo{author}{\bibfnamefont{K.}~\bibnamefont{Wiesenfeld}},
  \bibinfo{journal}{Phys. Rev. A}
  \textbf{\bibinfo{volume}{38}}(\bibinfo{number}{1}), \bibinfo{pages}{364}
  (\bibinfo{year}{1998}).

\bibitem{shape_memory}
\bibinfo{author}{\bibfnamefont{E.}~\bibnamefont{Vives}},
  \bibinfo{author}{\bibfnamefont{J.}~\bibnamefont{Ort{\'i}n}},
  \bibinfo{author}{\bibfnamefont{L.}~\bibnamefont{Ma{\~n}oso}},
  \bibinfo{author}{\bibfnamefont{I.}~\bibnamefont{R{\`a}fols}},
  \bibnamefont{and}
  \bibinfo{author}{\bibfnamefont{R.}~\bibnamefont{P{\'e}rez-Magran{\'e}}},
  \bibinfo{journal}{Phys. Rev. Lett.}
  \textbf{\bibinfo{volume}{72}}(\bibinfo{number}{11}), \bibinfo{pages}{1694}
  (\bibinfo{year}{1994}).

\bibitem{sethna1}
\bibinfo{author}{\bibfnamefont{J.~P.} \bibnamefont{Sethna}},
  \bibinfo{author}{\bibfnamefont{K.}~\bibnamefont{Dahmen}},
  \bibinfo{author}{\bibfnamefont{S.}~\bibnamefont{Kartha}},
  \bibinfo{author}{\bibfnamefont{J.~A.} \bibnamefont{Krumhansl}},
  \bibinfo{author}{\bibfnamefont{B.~W.} \bibnamefont{Roberts}},
  \bibnamefont{and} \bibinfo{author}{\bibfnamefont{J.~D.} \bibnamefont{Shore}},
  \bibinfo{journal}{Phys. Rev. Lett.}
  \textbf{\bibinfo{volume}{70}}(\bibinfo{number}{21}), \bibinfo{pages}{3347}
  (\bibinfo{year}{1993}).

\bibitem{sethna2}
\bibinfo{author}{\bibfnamefont{O.}~\bibnamefont{Perkovi{\'c}}},
  \bibinfo{author}{\bibfnamefont{K.~A.} \bibnamefont{Dahmen}},
  \bibnamefont{and} \bibinfo{author}{\bibfnamefont{J.~P.}
  \bibnamefont{Sethna}}, \bibinfo{journal}{Phys. Rev. Lett.}
  \textbf{\bibinfo{volume}{75}}(\bibinfo{number}{24}), \bibinfo{pages}{4528}
  (\bibinfo{year}{1995}).

\bibitem{sethna3}
\bibinfo{author}{\bibfnamefont{O.}~\bibnamefont{Perkovi{\'c}}},
  \bibinfo{author}{\bibfnamefont{K.~A.} \bibnamefont{Dahmen}},
  \bibnamefont{and} \bibinfo{author}{\bibfnamefont{J.~P.}
  \bibnamefont{Sethna}}, \emph{\bibinfo{title}{Disorder-induced critical
  phenomena in hysteresis: A numerical scaling analysis}},
  \bibinfo{note}{cond-mat \#9609072, Los Alamos Nat'l Laboratory, Los Alamos,
  N. M.; http://xxx.lanl.gov/abs/cond-mat/960972}.

\bibitem{sethna4}
\bibinfo{author}{\bibfnamefont{O.}~\bibnamefont{Perkovi{\'c}}},
  \bibinfo{author}{\bibfnamefont{K.~A.} \bibnamefont{Dahmen}},
  \bibnamefont{and} \bibinfo{author}{\bibfnamefont{J.~P.}
  \bibnamefont{Sethna}}, \bibinfo{journal}{Phys. Rev. B.}
  \textbf{\bibinfo{volume}{59}}(\bibinfo{number}{9}), \bibinfo{pages}{6106}
  (\bibinfo{year}{1999}).

\bibitem{sethna5}
\bibinfo{author}{\bibfnamefont{K.~A.} \bibnamefont{Dahmen}} \bibnamefont{and}
  \bibinfo{author}{\bibfnamefont{J.~P.} \bibnamefont{Sethna}},
  \bibinfo{journal}{Phys. Rev. B.}
  \textbf{\bibinfo{volume}{53}}(\bibinfo{number}{22}), \bibinfo{pages}{14872}
  (\bibinfo{year}{1996}).

\bibitem{dahmen}
\bibinfo{author}{\bibfnamefont{K.~A.} \bibnamefont{Dahmen}},
  \emph{\bibinfo{title}{Hysteresis, Avalanhces, and Disorder Induced Critical
  Scaling: A Renormalization Group Approach}}, Ph.D. thesis,
  \bibinfo{school}{Cornell University} (\bibinfo{year}{1995}).

\bibitem{urbach}
\bibinfo{author}{\bibfnamefont{J.~S.} \bibnamefont{Urbach}},
  \bibinfo{author}{\bibfnamefont{R.~C.} \bibnamefont{Madison}},
  \bibnamefont{and} \bibinfo{author}{\bibfnamefont{J.~T.}
  \bibnamefont{Markert}}, \bibinfo{journal}{Phys. Rev. Lett.}
  \textbf{\bibinfo{volume}{75}}(\bibinfo{number}{2}), \bibinfo{pages}{276}
  (\bibinfo{year}{1995}).

\bibitem{stanley1}
\bibinfo{author}{\bibfnamefont{S.}~\bibnamefont{Zapperi}},
  \bibinfo{author}{\bibfnamefont{P.}~\bibnamefont{Cizeau}},
  \bibinfo{author}{\bibfnamefont{G.}~\bibnamefont{Durin}}, \bibnamefont{and}
  \bibinfo{author}{\bibfnamefont{H.~E.} \bibnamefont{Stanley}},
  \bibinfo{journal}{Phys. Rev. B}
  \textbf{\bibinfo{volume}{58}}(\bibinfo{number}{10}), \bibinfo{pages}{6353}
  (\bibinfo{year}{1998}).

\bibitem{lieneweg}
\bibinfo{author}{\bibfnamefont{U.}~\bibnamefont{Lieneweg}} \bibnamefont{and}
  \bibinfo{author}{\bibfnamefont{W.}~\bibnamefont{Grosse-Nobis}},
  \bibinfo{journal}{Intern. J. Magnetism} \textbf{\bibinfo{volume}{3}},
  \bibinfo{pages}{11} (\bibinfo{year}{1972}).

\bibitem{stanley2}
\bibinfo{author}{\bibfnamefont{D.}~\bibnamefont{Spasojevi{\'c}}},
  \bibinfo{author}{\bibfnamefont{S.}~\bibnamefont{Bukvi{\'c}}},
  \bibinfo{author}{\bibfnamefont{S.}~\bibnamefont{Milosevi{\'c}}},
  \bibnamefont{and} \bibinfo{author}{\bibfnamefont{H.~E.}
  \bibnamefont{Stanley}}, \bibinfo{journal}{Phys. Rev. E}
  \textbf{\bibinfo{volume}{54}}(\bibinfo{number}{3}), \bibinfo{pages}{2531}
  (\bibinfo{year}{1996}).

\bibitem{kuntz}
\bibinfo{author}{\bibfnamefont{M.~C.} \bibnamefont{Kuntz}},
  \emph{\bibinfo{title}{Universality in {B}arkhausen noise: Simulation and
  experiment}}, \bibinfo{note}{unpublished}.

\bibitem{NotMeanField}
\bibinfo{note}{This is not the same as a mean-field model, because the model
  contains nearest neighbor interactions along with the infinite range
  interactions.}

\bibitem{narayan2}
\bibinfo{author}{\bibfnamefont{O.}~\bibnamefont{Narayan}},
  \bibinfo{journal}{Phys. Rev. Lett.}
  \textbf{\bibinfo{volume}{77}}(\bibinfo{number}{18}), \bibinfo{pages}{3855}
  (\bibinfo{year}{1996}).

\bibitem{robbins}
\bibinfo{author}{\bibfnamefont{H.}~\bibnamefont{Ji}} \bibnamefont{and}
  \bibinfo{author}{\bibfnamefont{M.~O.} \bibnamefont{Robbins}},
  \bibinfo{journal}{Phys. Rev. B}
  \textbf{\bibinfo{volume}{46}}(\bibinfo{number}{22}), \bibinfo{pages}{14519}
  (\bibinfo{year}{1992}).

\bibitem{BelowRcFrontProp}
\bibinfo{note}{The non-self-organized scaling described by Ji and Robbins also
  occurs below $R_c$ in the model of Sethna \protect\emph{et al.} with domain
  nucleation. Below $R_c$ with domain nucleation, the front sees some
  preflipped regions ahead which act as a second source of random, correlated
  disorder. However, below $R_c$ these regions typically have a small radius
  given by the correlation length $\xi\sim((R_c-R)/(R_c))^{-\nu}$. On a scale
  large compared to the correlation length the behavior is still well described
  by the front propagation model of Ji and Robbins.}

\bibitem{magni}
\bibinfo{author}{\bibfnamefont{A.}~\bibnamefont{Magni}},
  \bibinfo{journal}{Phys. Rev. B}
  \textbf{\bibinfo{volume}{59}}(\bibinfo{number}{2}), \bibinfo{pages}{985}
  (\bibinfo{year}{1999}).

\bibitem{petta}
\bibinfo{author}{\bibfnamefont{J.~R.} \bibnamefont{Petta}},
  \bibinfo{author}{\bibfnamefont{M.~B.} \bibnamefont{Weissman}},
  \bibnamefont{and} \bibinfo{author}{\bibfnamefont{G.}~\bibnamefont{Durin}},
  \bibinfo{journal}{Phys. Rev. E}
  \textbf{\bibinfo{volume}{57}}(\bibinfo{number}{6}), \bibinfo{pages}{6363}
  (\bibinfo{year}{1998}).

\bibitem{sigmanuz}
\bibinfo{note}{In earlier
  papers\protect\cite{sethna1,sethna2,sethna3,sethna4,sethna5}, our group has
  called the exponent relating avalanche size to avalanche duration $\sigma\nu
  z$. $\sigma$ is the exponent describing the growth of the cutoff in the
  avalanche size distribution with increasing disorder. $\nu$ is the exponent
  describing the growth of the correlation function as the critical disorder is
  approached. $z$ describes the dynamics of the model.}

\bibitem{invertedparabola}
\bibinfo{note}{This shape is very well fit by an inverted parabola. In mean
  field theory, this fit seems to be exact.}

\bibitem{CrossTerms}
\bibinfo{note}{There should also be contributions from cross-terms between
  avalanches, but this will contribute only at a time $\theta$ which grows as
  the field is ramped more and more slowly. In the limit of an adiabatically
  slowly increasing field, these cross terms will affect only the $\omega=0$
  scaling of the power spectrum. Here, we are calculating the scaling behavior
  for large $\omega$.}

\bibitem{cote1}
\bibinfo{author}{\bibfnamefont{P.~J.} \bibnamefont{Cote}} \bibnamefont{and}
  \bibinfo{author}{\bibfnamefont{L.~V.} \bibnamefont{Meisel}},
  \bibinfo{journal}{Phys. Rev. Lett.}
  \textbf{\bibinfo{volume}{67}}(\bibinfo{number}{10}), \bibinfo{pages}{1334}
  (\bibinfo{year}{1991}).

\bibitem{cote2}
\bibinfo{author}{\bibfnamefont{L.~V.} \bibnamefont{Meisel}} \bibnamefont{and}
  \bibinfo{author}{\bibfnamefont{P.~J.} \bibnamefont{Cote}},
  \bibinfo{journal}{Phys. Rev. B}
  \textbf{\bibinfo{volume}{46}}(\bibinfo{number}{17}), \bibinfo{pages}{10822}
  (\bibinfo{year}{1992}).

\bibitem{bertotti}
\bibinfo{author}{\bibfnamefont{G.}~\bibnamefont{Bertotti}},
  \bibinfo{author}{\bibfnamefont{F.}~\bibnamefont{Fiorillo}}, \bibnamefont{and}
  \bibinfo{author}{\bibfnamefont{A.}~\bibnamefont{Montorsi}},
  \bibinfo{journal}{J. Appl. Phys.}
  \textbf{\bibinfo{volume}{67}}(\bibinfo{number}{9}), \bibinfo{pages}{5574}
  (\bibinfo{year}{1990}).

\bibitem{jcf}
\bibinfo{author}{\bibfnamefont{H.~J.} \bibnamefont{Jensen}},
  \bibinfo{author}{\bibfnamefont{K.}~\bibnamefont{Christensen}},
  \bibnamefont{and} \bibinfo{author}{\bibfnamefont{H.~C.}
  \bibnamefont{Fogedby}}, \bibinfo{journal}{Phys. Rev. B}
  \textbf{\bibinfo{volume}{40}}(\bibinfo{number}{10}), \bibinfo{pages}{7425}
  (\bibinfo{year}{1989}).

\bibitem{NarayanPC}
\bibinfo{note}{Narayan uses the same method as Tang {\it et
  al.}\protect\cite{flux_line} use for flux-line depinning: their formula is
  valid in one dimension. Narayan's published formula\protect\cite{narayan2} is
  valid for dimension $d=2$: the dimension-independent formula in the text is
  due to him (Onuttom Narayan, private communication). Zapperi {\it et
  al.}\protect\cite{stanley1} do a similar computation for three dimensions,
  but appear to get a different result.}

\bibitem{narayan1}
\bibinfo{author}{\bibfnamefont{O.}~\bibnamefont{Narayan}} \bibnamefont{and}
  \bibinfo{author}{\bibfnamefont{D.~S.} \bibnamefont{Fisher}},
  \bibinfo{journal}{Phys. Rev. B}
  \textbf{\bibinfo{volume}{48}}(\bibinfo{number}{10}), \bibinfo{pages}{7030}
  (\bibinfo{year}{1993}).

\bibitem{Durin}
\bibinfo{author}{\bibfnamefont{G.}~\bibnamefont{Durin}},
  \bibinfo{author}{\bibfnamefont{G.}~\bibnamefont{Bertotti}}, \bibnamefont{and}
  \bibinfo{author}{\bibfnamefont{A.}~\bibnamefont{Magni}},
  \bibinfo{journal}{Fractals}
  \textbf{\bibinfo{volume}{3}}(\bibinfo{number}{2}), \bibinfo{pages}{351}
  (\bibinfo{year}{1995}).

\bibitem{ABBM1}
\bibinfo{author}{\bibfnamefont{B.}~\bibnamefont{Allesandro}},
  \bibinfo{author}{\bibfnamefont{C.}~\bibnamefont{Beatrice}},
  \bibinfo{author}{\bibfnamefont{G.}~\bibnamefont{Bertotti}}, \bibnamefont{and}
  \bibinfo{author}{\bibfnamefont{A.}~\bibnamefont{Montorsi}},
  \bibinfo{journal}{J. Appl. Phys.}
  \textbf{\bibinfo{volume}{68}}(\bibinfo{number}{6}), \bibinfo{pages}{2901}
  (\bibinfo{year}{1990}).

\bibitem{ABBM2}
\bibinfo{author}{\bibfnamefont{B.}~\bibnamefont{Alessandro}},
  \bibinfo{author}{\bibfnamefont{C.}~\bibnamefont{Beatrice}},
  \bibinfo{author}{\bibfnamefont{G.}~\bibnamefont{Bertotti}}, \bibnamefont{and}
  \bibinfo{author}{\bibfnamefont{A.}~\bibnamefont{Montorsi}},
  \bibinfo{journal}{J. Appl. Phys.}
  \textbf{\bibinfo{volume}{68}}(\bibinfo{number}{6}), \bibinfo{pages}{2908}
  (\bibinfo{year}{1990}).

\bibitem{flux_line}
\bibinfo{author}{\bibfnamefont{C.}~\bibnamefont{Tang}},
  \bibinfo{author}{\bibfnamefont{S.}~\bibnamefont{Feng}}, \bibnamefont{and}
  \bibinfo{author}{\bibfnamefont{L.}~\bibnamefont{Golubovic}},
  \bibinfo{journal}{Phys. Rev. Lett}
  \textbf{\bibinfo{volume}{72}}(\bibinfo{number}{8}), \bibinfo{pages}{1264}
  (\bibinfo{year}{1994}).

\end{thebibliography}
	\printtables
	\printfigures
\end{document}